%Paper: hep-th/9503086
%From: David Kutasov <kutasov@yukawa.uchicago.edu>
%Date: Mon, 13 Mar 1995 09:19:36 -0600

\input harvmac

\def\np#1#2#3{Nucl. Phys. {\bf B#1} (#2) #3}
\def\pl#1#2#3{Phys. Lett. {\bf #1B} (#2) #3}

\def\physrev#1#2#3{Phys. Rev. {\bf D#1} (#2) #3}

\def\prep#1#2#3{Phys. Rep. {\bf #1} (#2) #3}

\def\cmp#1#2#3{Comm. Math. Phys. {\bf #1} (#2) #3}

\def\tilde{\widetilde}

\Title{hep-th/9503086, EFI-95-11}
{\vbox{\centerline{A Comment on Duality in N=1 Supersymmetric}
\centerline{Non-Abelian Gauge Theories}}}
\bigskip
\centerline{D. Kutasov}
\vglue .5cm
\centerline{Enrico Fermi Institute and}
\centerline{Department of Physics}
\centerline{University of Chicago}
\centerline{Chicago, IL 60637, USA}
\vglue .3cm

\bigskip\bigskip

\noindent
Recently N. Seiberg has shown that certain $N=1$
supersymmetric Yang -- Mills theories
have the property that their infrared physics can be equivalently described
by {\it two different} gauge theories, with the strong coupling region
of one corresponding to the weak coupling region of the other. This duality
leads to interesting results about the infrared dynamics of these theories.
We generalize Seiberg's ideas to a new class of models for which
a similar duality allows one to obtain rather detailed dynamical
information about the long distance physics.

\Date{3/95}
%\draftmode

\nref\ads{I. Affleck, M. Dine, and N. Seiberg, \np{241}{1984}{493};
\np{256}{1985}{557}}%
\nref\nsvz{V.A. Novikov, M.A. Shifman, A. I.  Vainstein and V. I.
Zakharov, \np{223}{1983}{445}; \np{260}{1985}{157};
 \np{229}{1983}{381}}
\nref\svholo{M.A. Shifman and A.I Vainshtein, \np{277}{1986}{456};
\np{359}{1991}{571}}%
\nref\cern{D. Amati, K. Konishi, Y. Meurice, G.C. Rossi and G.
Veneziano, \prep{162}{1988}{169} and references therein}%
\nref\nonren{N. Seiberg, hep-ph/9309335, \pl{318}{1993}{469}}%
\nref\natii{N. Seiberg, hep-th/9402044, \physrev{49}{1994}{6857}}%
\nref\ils{K. Intriligator, R.G. Leigh and N. Seiberg, hep-th/9403198,
\physrev{50}{1994}{1092}
}%
\nref\sw{N. Seiberg and E. Witten, hep-th/9407087, \np{426}{1994}{19};
hep-th/9408099, \np{431} {1994} {484}
}%
\nref\intse{K. Intriligator and N. Seiberg, hep-th/9408155,
\np{431} {1994} {551}
}%
\nref\iss{K. Intriligator, N. Seiberg and S. Shenker, hep-ph/9410203,
\pl{342}{1995}{152}}%

Recently, important progress has been made in analyzing the
low energy properties of supersymmetric gauge theories in four dimensions
\refs{\ads-\iss}. Among the most remarkable is the observation by
N. Seiberg
\ref\nati{N. Seiberg, hep-th/9411149,
\np{435}{1995}{129}}\ that the infrared physics of
a large class of
Supersymmetric Yang Mills (SYM) theories, with $N_f$ flavors
of quarks in the fundamental representation of an $SU(N_c)$,
$SO(N_c)$ or $SP(N_c)$ gauge group is identical to that of another
SYM theory with an
in general  different number of colors $\tilde{N_c}$, but the same number
of flavors, $N_f$, as well as a set of gauge singlet fields. This duality
is very useful for  studying the long distance
properties of these models since it interchanges
strong and weak coupling. It has been used in \nati\ to achieve a
detailed understanding of the infrared degrees of freedom
of these theories in their non -- abelian Coulomb phase. Then, by turning
on masses one could also study some aspects of the dynamics
in a confining phase \nati.

The nature, physical origin and generality of this duality
remain, however, largely unclear, and it seems useful to construct
some additional
examples of this remarkable phenomenon. The purpose of this note is to
propose and briefly discuss a class of models that seem to be promising
candidates to exhibit a similar duality symmetry. Hopefully,
additional examples may shed more light on the dynamics of
strongly interacting gauge theories. We start with some
general remarks.

The $\beta$ function of $N=1$ SYM theories with matter fields $\Phi_i$
in representations $R_i$ of a gauge group $G$ is known exactly \svholo:
\eqn\bfun{\beta(\alpha)=-{\alpha^2\over2\pi}{3T(G)-\sum_iT(R_i)(1-\gamma_i)
\over
1-T(G){\alpha\over2\pi}}}
where $\alpha=g^2/4\pi$, $\gamma_i(\alpha)$ are the anomalous dimensions
of $\Phi_i$, given by:
\eqn\gfun{\gamma_i(\alpha)=-C_2(R_i){\alpha\over\pi}+O(\alpha^2)}
with
\eqn\trr{\eqalign{{\rm Tr}_R(T^aT^b)=&T(R)\delta^{ab}\cr
                             T^aT^a=&C_2(R) I\cr
                             T(G)\equiv&T(R={\rm adjoint}).\cr}}
Non -- trivial fixed points can arise if
\eqn\bo{3T(G)-\sum_i T(R_i)
(1-\gamma_i)=0.}
At the fixed points, $\gamma_i$ are related to the scaling
dimensions of
$\Phi_i$ through $\gamma_i+2=2d_i$, and the dimensions
$d_i$ are related to the $R$ charges of $\Phi_i$, $B_i$, by
the superconformal algebra,
$d_i={3\over2}B_i$. Therefore, we can write \bo\ as:
\eqn\anfr{T(G)+\sum_iT(R_i)(B_i-1)=0.}
Eq. \anfr\ is also the condition for an $R$ symmetry under which the
supercoordinates
$\theta_\alpha$ have charge one and $\Phi_i$ have charges $B_i$ to be
anomaly free. In general there may be many $R$ symmetries that satisfy
\anfr\ and thus are anomaly free. One of these becomes part of the
infrared (IR) superconformal algebra. It is interesting \bo, \anfr\ that
the $R$ symmetry which participates in the
IR superconformal algebra is actually an anomaly free symmetry of the
theory throughout the RG flow (in the absence of superpotentials).

To study the IR fixed point of SYM it is important to understand which of
the symmetries satisfying \anfr\ is the right one. There are some
situations in which the answer is known: e.g. if all representations $R_i$
are the same $R_i=R$, $i=1,\cdots, M$, the $R$ symmetry does not \nati\  break
the symmetry between the different representations, and
$B_i=B=1-{T(G)\over MT(R)}$. If the fixed point is perturbative
\ref\bankszaks{T. Banks and A. Zaks, \np{196}{1982}{189}},
i.e. $3T(G)-\sum T(R_i)<<T(G)$, the $B_i$ are close to their UV values,
$B_i\simeq2/3+\alpha\delta_i$ and one can use \gfun\ to compute $\delta_i$.
However, in general, it is not clear how to determine $B_i$ \anfr\ to
pick the right $R$ charge. Even more importantly, it is not clear in
general whether the theory ends up in the infrared in a non -- abelian
Coulomb phase or not (for examples of the issues involved, see \intse, \iss).

We will consider SYM with gauge group $G=SU(N_c)$ and matter
superfields $X$ in the adjoint representation of
the gauge group, $N_f$ multiplets $Q^i$ in the fundamental $(N_c)$
representation and $N_f$ multiplets $\tilde Q_{\tilde i}$ in the $\bar N_c$
$(i, \tilde i=1,\cdots, N_f)$. The theory is asymptotically free
for $N_f<2N_c$. It is natural to assign the same $R$ charge
$B_f$ to all the fundamental multiplets $Q^i$, $\tilde Q_{\tilde i}$, and a
different charge $B_a$
to the adjoint, $X$. The anomaly cancellation
condition \anfr\ then takes the form
\eqn\an{N_fB_f+N_cB_a=N_f}
and so the issue of which $R$ symmetry becomes part of the superconformal
algebra in the infrared arises. As we will see later,
duality will allow us to sidestep
this ambiguity. The attitude we will take to this model is heavily
influenced by \nati. We start by asking whether there could be a dual
picture of the dynamics. It is instructive to start the discussion with
the baryons, which in previous studies \nati\  revealed most directly the form
of
the duality transformation. In this theory, there are many baryon -- like
operators of the form\foot{We will mention additional baryon -- like
operators below.}
\eqn\baryons{ B^{[i_1,\cdots,i_k][i_{k+1}\cdots,i_{N_c}]}=
\epsilon^{\alpha_1,\cdots,\alpha_{N_c}}X_{\alpha_1}^{\beta_1}
X_{\alpha_2}^{\beta_2}\cdots X_{\alpha_k}^{\beta_k}Q^{i_1}_{\beta_1}
\cdots Q^{i_k}_{\beta_k}Q^{i_{k+1}}_{\alpha_{k+1}}\cdots
Q^{i_{N_c}}_{\alpha_{N_c}}}
$\alpha_i$, $\beta_j=1,\cdots, N_c$ are color indices. For given $k$ there are
${N_f \choose k}{N_f \choose N_c-k}$ operators
$B^{[i_1,\cdots,i_k][i_{k+1}\cdots,i_{N_c}]}$, so the total number of
baryon operators is
\eqn\smm{\sum_{k=0}^{N_c}{N_f \choose k}{N_f \choose N_c-k}=
{2N_f \choose N_c}.}
Note that only terms with $k\leq N_f$, $N_c-k\leq N_f$
contribute to the sum \smm.
Hence such operators exist only for $N_f\geq N_c/2$. We will restrict
ourselves to this region for the rest of the paper.

We see that the  spectrum of baryons may exhibit a symmetry under
$N_c\to 2N_f-N_c$ (with $N_f$ fixed). The precise mapping would be
\eqn\mapp{B_{\rm el}^{[i_1,\cdots,i_k][i_{k+1}\cdots,i_{N_c}]}\sim
B_{\rm mag}^{[j_1,\cdots,j_p][j_{p+1}\cdots,j_{2N_f-N_c}]}}
with $p=N_f-N_c+k$. We have denoted here the baryons \baryons\ by
$B_{\rm el}$ whereas $B_{\rm mag}$ are the baryon operators in the dual
(``magnetic'') theory discussed below.

Thus one is tempted to attempt to construct a theory dual to the one we
are studying, with $N_c\to\tilde N_c=2N_f-N_c$, such that the baryon
operators of the two theories are identified as described by \mapp.
Assuming the existence of such a duality, the identification \mapp\ leads to an
additional relation between the $R$ charges $B_f$ and $B_a$ and allows us
to fix them (as well as the $R$ charges in the dual theory)
unambiguously. One finds:
\eqn\rcharge{B_a={2\over3};\;\;\;B_f=1-{2\over3}{N_c\over N_f}.}
This leads to a puzzle. We find that the $R$ charge of the adjoint
multiplet $X$ is the same in the IR fixed point as in the UV one,
$B_a=2/3$. But in \bfun, \gfun\ we saw that even for perturbative fixed
points, i.e. when $2N_c-N_f<<N_c$, the dimension (and $R$ charge) of $X$
receives a non -- vanishing contribution at first order in $\alpha$, and
therefore is different in the IR and in the UV.
We believe that the resolution is that the theory under consideration has
a non -- zero superpotential \foot{For $N_c=2$ $W(X)=0$ and the
present discussion has to be reconsidered.}
\eqn\sp{W_{\rm el}(X)=c{\rm Tr} X^3.}
Without the superpotential \sp, the theory will presumably
flow in the infrared to a fixed point with $g=g_0^*$ and $c=0$,
which we will have nothing to say about here. With the superpotential \sp\
there is a possible  additional
fixed point in the $(g,c)$ plane
at $g=g_1^*$, $c=c^*\not=0$. Indeed, if
one first flows to $g\simeq g_0^*$ at $c=0$, the operator $X^3$ becomes
relevant and adding it to the action will send the theory to a new
(hopefully non -- trivial) fixed point. It is this fixed point that
is described by \rcharge.
Note that the superpotential
\sp\ respects only the $R$ symmetry \rcharge. Therefore, a way of
describing what we did is to say that we fixed the ambiguity
in determining the $R$ charge present in \an\ by
adding the superpotential \sp\ thereby eliminating the other $R$ symmetry.
Of course, we actually got \rcharge\ by requiring duality, which is
independent of the discussion of the superpotential.

To summarize the structure of the original, ``electric'', model
we give the transformation properties of the different operators
under the anomaly free global symmetry,
\eqn\globsym{SU(N_f)\times SU(N_f) \times U(1)_B \times U(1)_R }

\eqn\qtransl{\eqalign{
Q &\qquad (N_f,1,1, 1-{2\over3}{N_c\over N_f}) \cr
\tilde Q & \qquad (1, \overline N_f,-1,1-{2\over3}{N_c\over N_f})\cr
X &\qquad (1,1,0, {2\over3}) .\cr
}}

We are now ready to describe the dual theory, with gauge group
$SU(2N_f-N_c)$. In addition to the dual
quarks $q_i, \tilde q^{\tilde i}$ and the adjoint field (which we call $Y$
to distinguish it from $X$), we have to
add to the dual theory two gauge singlet chiral superfields,
\eqn\newf{\eqalign{M^i_{\tilde i}=Q^i\tilde Q_{\tilde i}\cr
                   N^i_{\tilde i}=Q^iX\tilde Q_{\tilde i}\cr}}
which exists in the original theory and can not be described
in terms of the dual variables $q, \tilde q, Y$. One does not need
to add operators of the form $QX^n\tilde Q$ with $n\geq2$ because
they do not belong to the chiral ring in
the presence of the superpotential\foot{The duality
discussed here and in \nati\ is probably restricted to the chiral rings
of the two dual models.}  \sp. The full list of fields in the
dual model with their transformation properties under
the global symmetry \globsym\ is:
\eqn\qnum{\eqalign{
q &\qquad (\overline N_f,1,{N_c\over2N_f-N_c}, 1-{2\over3}{2N_f-N_c\over N_f})
\cr
\tilde q & \qquad (1, N_f,-{N_c\over2N_f-N_c},1-{2\over3}{
2N_f-N_c\over N_f}) \cr
Y &\qquad (1,1,0, {2\over3}) \cr
M &\qquad (N_f,\overline N_f,0, 2-{4\over3}{N_c\over N_f}) \cr
N &\qquad (N_f,\overline N_f,0, {8\over3}-{4\over3}{N_c\over N_f}) .\cr
}}
Note that all the quantum numbers in \qnum\ are completely
determined by duality and the identifications \mapp, \newf.

Again following \nati, we note that one can (and should) add in the dual
model the
superpotential
\eqn\anw{W_{\rm mag}=M^i_{\tilde i}q_i Y\tilde  q^{\tilde i}+
             N^i_{\tilde i}q_i \tilde  q^{\tilde i}+{\rm Tr} Y^3.}
It is at first sight surprising to find an operator with UV dimension
four appear in \anw, especially since one is discussing IR properties
of the theory. Nevertheless, we will argue below that the first term in
$W_{\rm mag}$ is actually {\it not} always irrelevant (in both the technical
and colloquial sense).

An important test for the duality ansatz is the set of `t Hooft anomaly
matching conditions for the global symmetry group \globsym. Explicit
calculation shows that the anomaly matching conditions are satisfied; for
both \qtransl\ and \qnum\ we find the following anomalies:
\eqn\thooft{\eqalign{
SU(N_f)^3 \qquad &N_c d^{(3)}(N_f) \cr
SU(N_f)^2U(1)_R \qquad & -{2\over3}{N_c^2 \over N_f}d^{(2)}(N_f) \cr
SU(N_f)^2U(1)_B \qquad & N_cd^{(2)}(N_f) \cr
U(1)_R \qquad &-{2\over3}(N_c^2+1) \cr
U(1)_R^3 \qquad &{26\over 27}(N_c^2-1)-{16\over27}{N_c^4\over N_f^2} \cr
U(1)_B^2U(1)_R \qquad &-{4\over3}N_c^2 . \cr}}
Thus, one is inclined to believe that the duality conjecture may indeed be
valid. If one accepts this, one learns some interesting things about the
theory. Consider first the field $M^i_{\tilde i}$ \newf, \qnum. Its
scaling dimension ($3/2$ its $R$ charge) is:
\eqn\dimM{{\rm dim}M=3-{2N_c\over N_f}}
At $N_f\sim 2N_c$ it is $\sim2$ as appropriate for a weakly
coupled theory in the original electric variables \newf. As $N_f$
decreases, the coupling in the IR electric theory increases and the
dimension of $M$ \dimM\ decreases, until at $N_f=N_c$ it becomes one. At
that point general arguments
\ref\mack{G. Mack, \cmp{55}{1977}{1}}\ suggest that $M$ becomes a free
field. It is then natural to expect that the dimension of $M$ in
the IR remains one for $N_f<N_c$ as well.

{}From the point of view of the magnetic theory \qnum\ the behavior of $M$
is rather remarkable. Weak coupling
is at $N_f\simeq{2\over3}N_c$, and $M$ is a free field coupled to the
gauge sector by a
nonrenormalizable superpotential \anw. The coefficient of $MqY\tilde q$
in the superpotential decreases as we go to large distances,
and perturbatively (i.e. for $N_f$ not much larger than ${2\over3}N_c$)
it is clear that in the IR $M$ becomes a free field with
dimension one. However, as we increase $N_f$ (and with it the coupling
in the magnetic theory) the
anomalous dimension of $MqY\tilde q$ may become more and more negative
until, if the coupling
is strong enough, this irrelevant operator actually becomes relevant
and influences the IR
dynamics of $M$ and the strongly interacting magnetic gauge degrees of
freedom. Clearly, if this occurs it is due to a non -- perturbative effect
in the magnetic theory,
since the IR magnetic gauge coupling
must be larger than some critical coupling for the
coefficient of $MqY\tilde q$ in the
superpotential to become relevant. The original variables \qtransl\
can be used to study this magnetic strong coupling
effect at weak (electric) coupling.

One may think of the situation as follows\foot{
I thank N. Seiberg for a discussion of this point.}:
first consider the magnetic theory
without the $MqY\tilde q$ and $Nq\tilde q$ terms in $W_{\rm mag}$.
The model flows to a fixed point similar to the one described above for
the electric theory. The operators $q$, $\tilde q$, $Y$ have
dimensions related to the $R$ charges listed in \qnum. At the IR fixed
point of that theory one may study the operator $MqY\tilde q$.
If it is relevant at that
point, adding it to the superpotential will take us away from the fixed point
to a new one where $M$ is interacting. If it is irrelevant, the added
superpotential will not influence the IR dynamics, and $M$ will remain a free
field. An easy computation shows that the former occurs exactly when $N_f>N_c$
while the latter happens when $N_f\leq N_c$.

A similar discussion holds for
the operator $N$ and the related $Nq\tilde q$ interaction. One finds
that $N$ should be free when $N_f<{2\over3}N_c$, and interact non --
trivially otherwise. Indeed,
the operator $N^i_{\tilde i}$ has dimension $\qnum$,
\eqn\dimN{{\rm dim}N=4-{2N_c\over N_f}}
which goes to one at $N_f={2\over3}N_c$. For $N_f<{2\over3}N_c$
the magnetic gauge theory is not  asymptotically free, and hence the full
theory is free in the infrared.

The detailed picture of the two dual theories obtained above allows one to
study
the action of the duality map on other operators as well.
As an example, one may look at the baryonic operators\foot{
I thank A. Schwimmer for suggesting these
operators.},
\eqn\newbar{C^{[i_1,\cdots, i_k]}_{[j_1,\cdots, j_k]}=
\epsilon^{\alpha_1, \cdots,\alpha_{N_c}}
\epsilon_{\beta_1, \cdots,\beta_{N_c}}
Q^{i_1}_{\alpha_1}\cdots Q^{i_k}_{\alpha_k}
\tilde Q_{j_1}^{\beta_1}\cdots \tilde Q_{j_k}^{\beta_k}
X_{\alpha_{k+1}}^{\beta_{k+1}}\cdots
X_{\alpha_{N_c}}^{\beta_{N_c}}.}
For given $k$ there are ${N_f \choose k}^2$ such operators. Following
the same logic as before it is reasonable to assume that duality
relates here $Q^k\tilde Q^k X^{N_c-k}\leftrightarrow q^p\tilde q^p
Y^{2N_f-N_c-p}$ with $p=N_f-k$. Clearly, there is the same number
of both kinds of operators. One can then check using \qtransl,
\qnum\
that the global quantum numbers (in particular the $R$ charge and therefore
the scaling dimension at the IR fixed point) also agree.

The general structure of this set of theories is thus the following:
the theory is free when $N_f>2N_c$ and when $N_f<{2\over3}N_c$. The
former corresponds to free electric variables, with the fields $M$, $N$
corresponding to quark composites with dimensions two and three, repectively;
the latter corresponds to free
magnetic variables with
$M$ and $N$ being free fields with dimension one.  The theory is interacting
for
${2\over3}N_c<N_f<2N_c$. The model with $N_f=N_c$ is self dual
under the duality $N_c\to 2N_f-N_c$, which exchanges the region $N_c<N_f<2N_c$
with ${2\over3}N_c<N_f<N_c$. The field $M$ is free (with dimension one) when
$N_f<N_c$ but the full theory is not.

The next step should be to analyze the role of masses and flat directions
in the two dual theories. It is not difficult to study the decoupling of
quark flavors. Consider e.g. giving a mass to $Q^{N_f}$, $\tilde
Q_{\tilde N_f}$. In the electric variables this leads in the infrared to
decoupling of $Q^{N_f}$ and takes the theory from $N_f$ to $N_f-1$
flavors. In the dual magnetic theory with gauge group $SU(\tilde
N_c=2N_f-N_c)$ we should find that $\tilde N_c$  decreases by two units
as well as $N_f\to N_f-1$. This can indeed be checked;
the superpotential with the mass term is
\eqn\ww{W_{\rm mag}=M^i_{\tilde i}q_i Y\tilde q^{\tilde i}+N^i_{\tilde i}
\tilde q^{\tilde i}q_i+mM^{N_f}_{\tilde N_f}+cY^3.}
We can integrate out the massive fields, whose equations of motion lead
(among other things) to:
\eqn\eqmo{\eqalign{
q_{N_f} Y\tilde q^{\tilde N_f} &= -m  \cr
q_{N_f} \tilde q^{\tilde N_f} &= 0  \cr
q_{i} \tilde q^{\tilde N_f} &= 0  \cr
q_{N_f} \tilde q^{\tilde i} &= 0. \cr}}
This means that the gauge group is broken to $SU(2N_f-N_c-2)$
and the number of flavors decreases by one, the correct answer.
Note that here too it was important to keep the nonrenormalizable first
term in the superpotential \ww\ to get the right structure as masses are
turned on. As in \nati, adding the mass term to the Lagrangian of the electric
theory makes it more strongly interacting and eventually confining. In the
dual magnetic variables it corresponds \eqmo\ to
Higgsing the gauge group and thus
makes the theory more weakly coupled.

To conclude, we have seen again the power and beauty of the
duality of \nati; very few qualitative assumptions
led us to a rather detailed, and non -- trivial
to independently verify quantitative
picture of the infrared dynamics of a large class
of strongly interacting supersymmetric gauge theories.
Many questions remain. One would like to understand what (if any) non --
perturbative superpotential the theory can generate, along the lines of
\natii. It is important to understand the structure of the moduli space
of vacua (flat directions) and the relation to the $N=2$ SYM models of
\sw,
study the dynamics for $N_f\leq{N_c\over2}+1$, $N_c=2$, and try
to generalize to other fixed points of this theory and
to other gauge groups. More ambitiously, one would like
to develop a general framework for treating supersymmetric gauge
theories that makes duality manifest. For this it would be useful
to find additional examples of theories exhibiting similar symmetries.
All these and other issues are left for future work.

\bigskip

\centerline{{\bf Acknowledgments}}

I would like to thank J. Harvey, E. Martinec, A. Schwimmer
and N. Seiberg for discussions.
This work was supported in part by a DOE OJI award.

\listrefs

\end